\documentclass[english,prd,twocolumn,superscriptaddress,nofootinbib,preprintnumbers]{revtex4}
\usepackage[latin1]{inputenc}
\usepackage{graphicx}
\usepackage{amssymb}
\usepackage{color}
\usepackage{float}
\usepackage{amsmath}
\usepackage{amsfonts}
\usepackage{dcolumn}
\usepackage{hyperref}
\usepackage{subfigure}

\begin{document}
\title{Accretion onto a black hole in a string cloud background}

\author{Apratim Ganguly}\email{ganguly@ukzn.ac.za}
\affiliation{Astrophysics and Cosmology Research Unit,
 School of Mathematics, Statistics and Computer Science,
 University of KwaZulu-Natal, Private Bag X54001,
 Durban 4000, South Africa}

\author{Sushant G. Ghosh}\email{sghosh2@jmi.ac.in}
\affiliation{Astrophysics and Cosmology Research Unit,
 School of Mathematics, Statistics and Computer Science,
 University of KwaZulu-Natal, Private Bag X54001,
 Durban 4000, South Africa}
 \affiliation{Centre for Theoretical Physics,
 Jamia Millia Islamia,  New Delhi 110025
 India}

\author{Sunil D. Maharaj}\email{maharaj@ukzn.ac.za}
\affiliation{Astrophysics and Cosmology Research Unit,
 School of Mathematics, Statistics and Computer Science,
 University of KwaZulu-Natal, Private Bag X54001,
 Durban 4000, South Africa}

\begin{abstract}

We examine the accretion process onto the black hole with a string cloud background, where
the horizon of the  black hole has an enlarged radius $r_H=2 M/(1-\alpha)$, due to the string cloud parameter $\alpha\; (0 \leq \alpha < 1)$.
 The problem of stationary, spherically symmetric accretion of a polytropic fluid  is analysed to obtain
 an analytic solution for such a perturbation. Generalised expressions  for the
  accretion rate $\dot{M}$, critical radius $r_s$, and
 other flow parameters are found. The
accretion rate $\dot{M}$ is an explicit function of the black hole
mass $M$, as well as the gas boundary conditions and the string
cloud parameter $\alpha$. We also find the gas compression
ratios and temperature profiles below the accretion radius and at
the event horizon.  It is shown that the mass accretion rate, for
both the relativistic and the non-relativistic fluid by a black hole in
the string cloud model, increases with increase in $\alpha$.
\end{abstract}

\maketitle

\section{Introduction}

Black holes are amongst the most striking predictions of Einstein's
theory of general relativity. One of the most important effects of the black hole is its
tendency to accrete, and hence several aspects of the  spherical
accretion onto the black hole have been actively investigated in detail
over the last four decades (see \cite{Chakrabarti:1996cc} for a review).
Accretion is the term used by astrophysicists to describe the inflow
of matter towards a central gravitating object or towards the center
of the mass of an extended system.  It
may be pointed that accretion of matter onto black holes is one of
the most promising ideas  explaining  the highly luminous
active galactic nuclei (AGNs) and quasars.  The first study of spherical
accretion onto compact objects dates back more than forty years in
the seminal paper due to Bondi \cite{Bondi}. In this classic work,
the hydrodynamics of polytropic flow is studied within the Newtonian
framework, and it is found that either a settling or transonic
solution is mathematically possible for the gas accreting onto
compact objects. Note that the accretion rate is highest for the
transonic solution. The relativistic version of the same problem was
solved by Michel \cite{Michel} twenty years later.  Michel
\cite{Michel} investigated the steady state spherically symmetric
flow of a test gas onto a Schwarzschild black hole in the framework
of general relativity.  He showed that accretion onto the black hole
should be transonic. Michel's relativistic results attracted several
researchers \cite{keylist,stbook}. Spherical accretion and winds in
the context of general relativity have also been analyzed using equations of
state other than the polytrope. Other extensive studies
include the calculation of the frequency and luminosity spectra
\cite{shap73a}, the influence of an interstellar magnetic field in ionized gases \cite{shap73b}, and the changes
in accreting
processes when the black hole rotates \cite{shap74}. Several
radiative processes have been included by Blumenthal  and
Mathews \cite{blum}, and Brinkmann \cite{brink}. In addition
Malec \cite{malec} considered general relativistic
spherical accretion with and without back-reaction, and showed that
relativistic effects increase mass accretion when back-reaction is
absent. Accretion of a perfect fluid with a general equation of
state onto a Schwarzschild black hole has been investigated in
\cite{perfect, perfect1}, and a similar analysis for a charged black
hole has been done in \cite{charge1}. Accretion processes related  to  a charged black
hole were analyzed in \cite{Michel} and investigated further in
\cite{charge,Jamil,Sharif11,Sharif12}. The main aim of these studies
is to obtain the net energy output emitted by infalling gas with
application of black hole accretion to several classes of
astrophysical sources. It is understood that accretion onto a
black hole might be an important source of radiant energy. This may be
related to the accretion rate $\dot{M}$, and we may expect that an
increase  in $\dot{M}$ should lead to an increase in the luminosity
\cite{shap73a}.  In this paper, we consider the steady state spherical accretion onto
a black hole that has a cloud of strings in the background. We thereby
generalize the previous work of Michel \cite{Michel}. It may be
noted that  the study of Einstein's equations coupled with a
\textit{string cloud} may be very important as the relativistic
strings at a classical level can be used to construct applicable models
\cite{Le1}. Also, the universe can be represented as a collection of
extended (nonpoint) objects and 1-dimensional strings are the most
popular candidate for such fundamental objects. Hence the study of
the gravitational effects of matter, in the form of clouds of both
cosmic and fundamental strings, has  generated considerable
attention \cite{Synge}.  

Cosmic strings are a generic outcome of symmetry-breaking phase transitions in the early 
universe \cite{Kibble}, and further motivation comes from a potential role in large scale structure formation \cite{Vilenkin}.  Strings may
have been present in the early universe, and they play a role in the
seeding of density inhomogeneities \cite{dn}.  The magnitude of such strings are determined by the dimensionless parameter
\[ \frac{G\mu}{c^2} = \left(\frac{\eta}{m_{Pl}}\right)^2, \]
where $\eta$ is the energy scale of string and $m_{Pl}=\sqrt{hc/G}$ is the Planck mass.  
For the Nambu-Goto string model, using the Planck data, it has been
 shown that  a constraint on the string tension of $\frac{G\mu}{c^2} < 1.5 \times 10^{-7}$ at $95$ percent confidence that can be 
 improved to $\frac{G\mu}{c^2} < 1.3 \times 10^{-7}$   on inclusion of high-$ l $ CMB data \cite{Planks}.

It may be also pointed out that strings have become a very
important ingredient in many physical theories, and  the idea of
strings is fundamental in superstring theories \cite{as}.   The apparent
relationship between counting string states and the entropy of the
black hole horizon \cite{fl,Strominger:1996sh} suggests an association of
strings with black holes. Furthermore the intense level of activity in
string theory has lead to the idea that many of the classic vacuum
scenarios, such as the static Schwarzschild point black hole, may
have atmospheres composed of a fluid or field of strings \cite{pv}.
Many authors have found exact black hole solutions  with string
cloud backgrounds, for instance, in general relativity \cite{Le1,sgr},
in Einstein-Gauss-Bonnet models \cite{sgb}, and in Lovelock gravity
\cite{sll}, thereby generalizing the pioneering work of Letelier
\cite{Le1} who modified the Schwarzschild black hole for the string cloud
model. Glass and Krisch \cite{gk}  pointed out that allowing the
Schwarzschild mass parameter to be a function of radial position
creates an atmosphere with a string fluid stress-energy tensor around a
static, spherically symmetric object.

Interestingly, it turns out that the mass accretion rate $\dot{M}$
increases with the string cloud as a background in comparison to the
standard black hole. Also note that the mass accretion rate is
affected by the presence of higher dimensions \cite{John}.

The paper is organized as follows: In Sec. II we review the action, the energy momentum tensor
for a cloud of strings,  and the corresponding black hole solution. In Sec. III the
analytic general relativistic accretion
onto  a Schwarzschild black hole is appropriately generalized
to model spherical steady state accretion onto a black hole surrounded
by a cloud of strings. We calculate how the presence of a string cloud
would affect the mass accretion rate $\dot{M}$ of a gas onto a black
hole.   We also determine analytic  corrections to the critical radius,
the critical fluid velocity and the sound speed, and subsequently to  the mass accretion rate.
We then obtain expressions for the asymptotic behavior of the fluid density and the temperature
near the event horizon in Sect. IV. Finally we conclude in Sec. V.

 We use the following values for the physical constants for numerical computations and plots:
 $c = 3.00 \times 10^{10} \mathsf{cm. s}^{-1}$,
 $G = 6.674 \times 10^{-8} \mathsf{cm}^{3}. \mathsf{g.}^{-1} \mathsf{s}^{-2}$,
 $k_B = 1.380 \times 10^{-16} \mathsf{erg. K}^{-1}$,
 $M = M_{\odot} = 1.989 \times 10^{33} \mathsf{g}$,
 $m_b = m_{p} =1.67 \times 10^{-24} \mathsf{g}$,
 $n_{\infty} = 1 \mathsf{cm}^{-3}$,
 $T_{\infty} = 10^4 \mathsf{K}$.

\section{Schwarzschild black hole in a string cloud background}

In this model, we assume a spherically symmetric metric, and
steady state flow onto a nonrotating black hole of mass $M$ at rest.
 Recall that the
non-relativistic model was discussed by Bondi \cite{Bondi}, and the
standard 4-dimensional general relativistic version was developed by
Michel \cite{Michel}. The known analytic relativistic accretion
solution onto the Schwarzschild black hole by Michel \cite{Michel}
is generalized by considering a cloud of strings in the background.
To achieve this, we first briefly review  the theory on a cloud of
strings (see \cite{Le1} for further details) and the corresponding
modified Schwarzschild black hole.

The Nambu-Goto action of a string evolving in spacetime is given by
\[    I_{\mathcal{S}} = \int_{\Sigma} \mathcal{L} \;  d\lambda^{0} d\lambda^{1}, \hspace{0.2in} \mathcal{L} = m (\Gamma)^{-1/2} ,    \]
where $m$ is a positive constant that characterizes each string,
$(\lambda^{0}, \lambda^{1})$ is a parametrization of the world sheet
$\Sigma$ with $\lambda^{0}$ and $\lambda^{1}$ being timelike and
spacelike parameters \cite{Synge}, and $\Gamma$ is the determinant of
the induced metric on the string world sheet $\Sigma$   given by
\begin{equation}
 \Gamma_{a b} = g_{\mu \nu} \frac{\partial x^{\mu}}{\partial \lambda^{a}} \frac{\partial x^{\nu}}{\partial \lambda^{b}},
\end{equation}
and $\Gamma $ = det $\Gamma_{a b}$. Associated with the string worldsheet we have the bivector of the form
\begin{equation}
\label{eq:bivector}
     \Sigma^{\mu \nu} = \epsilon^{a b} \frac{\partial x^{\mu}}{\partial \lambda^{a}} \frac{\partial x^{\nu}}{\partial \lambda^{b}},
\end{equation}
where $\epsilon^{a b}$ denotes the two-dimensional Levi-Civita tensor given by $\epsilon^{0 1} = - \epsilon^{1 0} = 1$.
Within this setup, the Lagrangian density becomes
\[     \mathcal{L} = m \left[-\frac{1}{2} \Sigma^{\mu \nu} \Sigma_{\mu \nu}\right]^{1/2}.     \]
Further, since $T^{\mu \nu} = 2 \partial \mathcal{L}/\partial g^{\mu \nu}$, we obtain the energy momentum tensor for one string as
\begin{equation}
T^{\mu \nu} = m \Sigma^{\mu \rho} \Sigma_{\rho}^{\phantom{\rho} \nu}/(-\Gamma)^{1/2}.
\end{equation}
Hence, the energy momentum tensor for a cloud of string is
\begin{equation}
T^{\mu \nu} = \rho {\Sigma^{\mu \sigma} \Sigma_{\sigma}^{\phantom{\sigma} \nu}}/{(-\Gamma)^{1/2}  },
\end{equation}
where $\rho$ is the proper density of a string cloud. The quantity
$\rho \; (\Gamma)^{-1/2} $ is the gauge invariant quantity  called
the gauge-invariant density.

The general solution of  Einstein's equations for a string cloud
in 4-dimensions takes the form
\begin{eqnarray}
\label{stringmetric}
ds^2 &=& - \left(1- \frac{2M}{r}-\alpha \right) dt^2 + \left(1- \frac{2M}{r}-\alpha \right)^{-1} dr^2 \nonumber \\
&&  +r^{2}(d\theta^{2}+\sin^{2}\theta d\phi^{2}),
\end{eqnarray}
where we have set $G = c = 1 $ in this paper. Here $M$ arises as an
integration constant which is identified as the black hole mass  and
is not a function of  $\alpha$. The event horizon for the metric
(\ref{stringmetric})  has radius
\begin{equation}
 \label{horizon}
 r_H=\frac{2M}{1-\alpha}, \hspace{0.5in} \alpha \neq 1.
\end{equation}
In the limit $\alpha \rightarrow 0$, we recover the Schwarzschild
radius, and  close to unity the event horizon radius tends to
infinity. 
In general the string cloud parameter $\alpha \neq
1$.  We note that the case of static spherical  symmetry restricts
the value of the gauge-invariant density to $\rho (-\Gamma)^{1/2}  =
\alpha/r^2$ \cite{Le1}, and thereby $\alpha$ is a positive constant.
However, for the realistic model under consideration here the
string cloud parameter is restricted to $0 < \alpha < 1$. On the other hand, the
cloud of strings alone ($M = 0$) does not have a horizon; it
generates only a naked singularity at $ r = 0 $. This solution was
first obtained by Letelier \cite{Le1} and the metric represents the
black hole spacetime associated with a spherical mass $M$ centered
at the origin of the system of coordinates, surrounded by a
spherical cloud of strings. Furthermore it can be interpreted as the
metric associated with a global monopole.  In the string cloud
background, the Schwarzschild radius of the black hole is displaced
by the factor $(1-\alpha)^{-1}$.

\section{General equations for spherical accretion}

We now present the basic relations in spherical symmetry with
accreting matter, and describe the flow of
gas into the modified Schwarzschild black hole (\ref{stringmetric}). Also we
probe how the string cloud background affects the accretion rate
$\dot{M}$, the asymptotic compression ratio, and the temperature
profiles. We consider the steady state radial inflow of gas onto a
central mass $M$  by following the approach of Michel
\cite{Michel} and Shapiro \cite{stbook}. The gas is approximated as
a perfect fluid described by the energy momentum tensor
\begin{equation}
\label{energy-mom}
T^{\mu\nu} = \left( \rho + p \right)u^\mu u^\nu + p g^{\mu\nu},
\end{equation}
where  $\rho$ and $p$ are the fluid proper energy density and pressure respectively, and
\begin{equation}
\label{velocity}
u^\mu = \frac{d x^\mu }{ds},
\end{equation}
is the fluid 4-velocity  which obeys the normalization condition
$u^\mu u_ \mu = -1.$  We also define the proper baryon number
density  $n$, and the  baryon number flux $J^\mu  = n u^\mu$. All
these quantities are measured in the local inertial rest frame of
the fluid. The spacetime curvature is dominated by the compact
object and we ignore the self-gravity of the fluid. The accretion
process is based on two important conservation laws. Firstly, if no
particles are created or destroyed then particle number is conserved
and
\begin{equation}
\label{baryoncons}
\nabla_\mu J^\mu = \nabla_\mu ( n u^\mu ) = 0.
\end{equation}
Secondly, the conservation law is that of energy momentum which is governed by
\begin{equation}
\label{momencons} \nabla_\mu T^{\mu}_{\nu} = 0.
\end{equation}
The non-null components of the 4-velocity are $u^0 = dt/ds$ and $v(r) = u^1 = dr/ds $. Since $u_\mu u^\mu = -1$, and the velocity components vanish for $\mu > 1$ , we have
\begin{equation}
\label{zerocomp} u^0 = \left[\frac{v^2 + 1 -
\frac{2M}{r}-\alpha}{\left( 1 - \frac{2M}{r}-\alpha
\right)^2}\right]^{1/2}.
\end{equation}
Equation (\ref{baryoncons}) can be written as
\begin{equation}
\label{baryoncons1}
\frac{1}{r^2} \frac{d}{dr} \left( r^2 n v \right)= 0.
\end{equation}
Our assumptions of spherical  symmetry and steady state flow make (\ref{momencons}) comparatively easier to tackle. The $\nu = 0$ component is
\begin{equation}
\label{momencons0} \frac{1}{r^2} \frac{d}{dr} \left[ r^2 (\rho + p)
v \left(  1 - \frac{2M}{r}-\alpha +   v^2 \right)^{1/2} \right] = 0.
\end{equation}
The $\nu=1$ component can be simplified to
\begin{equation}
\label{momencons1}
v \frac{dv}{dr} = - \frac{dp}{dr} \left( \frac{ 1 - \frac{2M}{r}-\alpha + v^2  }{\rho + p} \right) -\frac{M}{r^2}.
\end{equation}
The above equations are generalization of the results obtained for
the standard Schwarzschild black hole \cite{Michel, stbook}.

\subsection{Accretion onto a black hole}
The accretion of matter onto black holes remains a classic problem
of contemporary astrophysics, as it does on the related problems of
active galactic nuclei and quasars, the mechanism of jets, and the
nature of certain galactic x-ray source.  Let us consider  spherical
steady state accretion onto a Schwarzschild black hole of mass $M$
in a string cloud background to obtain the mass accretion rate from
a qualitative analysis of (\ref{baryoncons1}) and
(\ref{momencons1}). For an adiabatic fluid there is no entropy
production and the conservation of mass-energy is governed by
\begin{equation}
\label{secondlaw}
T ds = 0 = d \left( \frac{\rho}{n}\right) + p~d\left(\frac{1}{n}\right),
\end{equation}
which may be put in the form
\begin{equation}
\label{slaw1}
\frac{d\rho}{dn} = \frac{\rho + p}{n}.
\end{equation}
We define the adiabatic sound speed $a$   via \cite{stbook}
\begin{equation}
\label{soundspd}
a^2 \equiv \frac{dp}{d\rho} = \frac{dp}{dn} \frac{n}{\rho + p},
\end{equation}
and we have used equation (\ref{slaw1}).
Using (\ref{soundspd}), the baryon and momentum conservation equations can be written as
\begin{align}
\label{bcon}
\frac{v'}{v} + \frac{n'}{n}+\frac{2}{r} &= 0, \\
\label{mcon1}
vv' + a^{2}\left( 1 - \frac{2M}{r}-\alpha + v^2\right) \frac{n'}{n}+\frac{M}{r^2} &= 0,
\end{align}
with $p'=(dp/dn)n'$ where  a dash ($'$) denotes a derivative with respect to $r$. With the help of the above equations, we obtain the system
\begin{align}
\label{system}
v' &= \frac{N_1}{N}, \nonumber \\
n' &= -\frac{N_2}{N},
\end{align}
where
\begin{subequations}\label{def}
\begin{align}
\label{def:1}
N_1 & = \frac{1}{n} \left[ \left(1 - \frac{2M}{r}-\alpha + v^2 \right)\frac{2a^2}{r} - \frac{M}{r^2}  \right], \\
\label{def:2}
N_2 & = \frac{1}{v} \left(\frac{2v^2}{r} - \frac{M}{r^2}  \right), \\
\label{def:3} N &= \frac{v^2 - \left( 1 - \frac{2M}{r}-\alpha + v^2
\right)a^2 }{vn}.
\end{align}
\end{subequations}
In the stationary accretion of gas onto the black hole, the amount of infalling
matter   per unit time $\dot{M}$, and other parameters are
determined by the gas properties and the gravitational field at large
distances.  For large $r$, the flow is subsonic i.e. $v < a$ and since
the sound speed must be subluminal, i.e., $a < 1$, we have $v^2 \ll
1$. The denominator (\ref{def:3}) is therefore
\begin{equation}
N \approx \frac{v^2 - a^2(1-\alpha)}{vn},
\end{equation}
and so $N < 0$ as $r \rightarrow \infty $ if we demand $v^2 <
a^2(1-\alpha)$. At the event horizon $r_H = 2M/(1-\alpha)$, and we have
\begin{equation}
N = \frac{v(1-a^2)}{n}.
\end{equation}
Under the causality constraint $a^2<1$,  we have $N >0$.  Therefore
$N$ should pass through a critical point $r_s$ where it goes to
zero. As the flow is assumed to be smooth everywhere, so  $N_{1}
~\text{and} ~N_{2}$ should also vanish at $r_s$, i.e.,  to avoid
discontinuities in the flow, we must have $N = N_{1} = N_{2} =0$ at
the radius $r_s$. This is nothing but the so-called {\it sonic
condition}. Hence, the flow must pass through a critical point
outside the event horizon, i.e., $r_H < r_s < \infty$. At the critical
point the system (\ref{def}) satisfies the condition
\begin{equation}
\label{sonic} v_{s}^{2} = \frac{a_{s}^{2}(1-\alpha)}{1 + 3a_{s}^{2}}
= \frac{M}{2r_{s}},
\end{equation}
where $v_s \equiv v(r_s)$ and $a_s \equiv a(r_s)$.  The quantities
with a subscript $s$ are defined at the critical point or the sonic
points of  the flow.  It can be clearly seen that the critical velocity
in this model is modified by the factor $(1-\alpha)$, and
the physically acceptable solution $v_{s}^{2}> 0 $ is ensured since
$0\leq \alpha <0$.

To calculate the mass accretion rate, we integrate
(\ref{baryoncons1}) over a 4-dimensional volume and multiply
by $m_b$, the mass of each baryon, to obtain
\begin{equation}
\label{accrate} \dot{M} = 4 \pi r^2 m_{b}nv,
\end{equation}
where $\dot{M}$ is an integration constant, independent of $r$,
having dimensions of mass per unit time. It is similar to the
Schwarzschild case.  Equations (\ref{baryoncons1}) and
(\ref{momencons0}) can be combined to yield
\begin{equation}
\label{bernoulli} \left( \frac{\rho + p}{n} \right)^2 \left( 1 -
\frac{2M}{r}-\alpha + v^2 \right) = \left( \frac{\rho_{\infty} +
p_{\infty}}{n_{\infty}}\right)^2,
\end{equation}
which is the modified relativistic Bernoulli equation for the
steady state accretion onto black holes surrounded by a cloud of
strings. Equations (\ref{accrate}) and (\ref{bernoulli}) are the basic
equations that characterize accretion onto a black hole with parameter $\alpha$
 where we have ignored the back-reaction of matter. In the
limit $\alpha=0$, our results reduce to those obtained in
\cite{Michel,stbook} for the standard Schwarzschild black hole.

\subsection{The polytropic solution}

In order to calculate   $\dot{M}$ explicitly and  all  the
fundamental characteristics of the flow, (\ref{accrate})
and (\ref{bernoulli}) must be supplemented with an equation of
state which is a  relation that characterizes the state of matter
of the gas. Following Bondi \cite{Bondi} and Michel \cite{Michel}, we
introduce a polytropic equation of state
\begin{equation}
\label{eos} p = K n^{\gamma},
\end{equation}
where $K$ and the adiabatic index $\gamma$ are constants. On
inserting (\ref{eos}) into the energy equation (\ref{secondlaw}) and
integrating, we obtain
\begin{equation}
\label{rho} \rho = \frac{K}{\gamma -1}n^{\gamma} + m_{b}n,
\end{equation}
where $m_{b}$ is an  integration constant obtained by matching with the
total energy density equation $\rho=m_{b}n+U$, where $m_{b}n$ is the
rest-mass energy density of the baryons and $U$ is the internal energy density.
Equations (\ref{eos}) and (\ref{rho}) give
\begin{equation}
 \label{inter}
 \gamma Kn^{\gamma-1}=\frac{a^2m_{b}}{(1-\frac{a^2}{\gamma-1})}.
\end{equation}
Using (\ref{rho}) and (\ref{inter}) we  can easily rewrite the
Bernoulli equation (\ref{bernoulli}) as
\begin{align}
\label{ber1}
&\left(1 + \frac{a^{2}}{\gamma - 1 - a^{2} }\right)^2 \left( 1  - \frac{2M}{r}-\alpha + v^2 \right) \nonumber \\
& = \left(1 +  \frac{a_{\infty}^{2}}{\gamma - 1 -
a_{\infty}^{2}}\right)^2.
\end{align}
At the critical radius $r_s$, using the relation (\ref{sonic}) and
inverting the above equation, we get
\begin{equation}
\label{berson} \left(1 + 3a_{s}^{2} \right) \left( 1 -
\frac{a_{s}^{2}}{\gamma -1} \right)^2 = \left( 1 -
\frac{a_{\infty}^{2}}{\gamma - 1} \right)^2.
\end{equation}
It must be noted that,  in general, the Bernoulli equation is
modified due to a string cloud background. However at the critical
radius, the form remains unchanged from the Schwarzschild case
\cite{stbook}.

For large but finite values of  $r$, i.e. $r \geq r_s$ the baryons will
be non-relativistic, i.e., $T \ll mc^2/k = 10^{13} K$ for
neutral hydrogen. In this regime we should have $a \leq a_s \ll 1$.
Expanding (\ref{berson}) up to second order in $a_s$ and
$a_{\infty}$, we obtain
\begin{align}
\label{sonsound}
a_{s}^{2} &\approx \frac{2}{5-3\gamma} a_{\infty}^{2}, \quad \gamma\neq\frac{5}{3}, \nonumber \\
&\approx \frac{2}{3}a_{\infty}, \quad \quad \quad ~\gamma =
\frac{5}{3}.
\end{align}
We thus obtain the critical radius $r_s$ in terms of the black hole
mass $M$ and the boundary condition $a_{\infty}$ from  (\ref{sonic}) and (\ref{sonsound}):
\begin{align}
\label{sonrad}
r_{s} &\approx \frac{5-3\gamma}{4}\frac{M}{a_{\infty}^2(1-\alpha)}, \quad \gamma\neq\frac{5}{3} \nonumber \\
&\approx \frac{3}{4} \frac{M}{ a_{\infty}(1-\alpha)}, \quad \quad \quad ~\gamma = \frac{5}{3}.
\end{align}
Also, for $a^{2}/(\gamma-1) \ll 1$, we get from (\ref{inter})
\begin{equation}
\label{barden1} \frac{n}{n_{\infty}} \approx \left(
\frac{a}{a_{\infty}} \right)^{2/(\gamma-1)}.
\end{equation}

\begin{widetext}

\begin{figure}
\begin{tabular}{|c|c|c|}
\hline
\includegraphics[width=7.5 cm, height=6 cm]{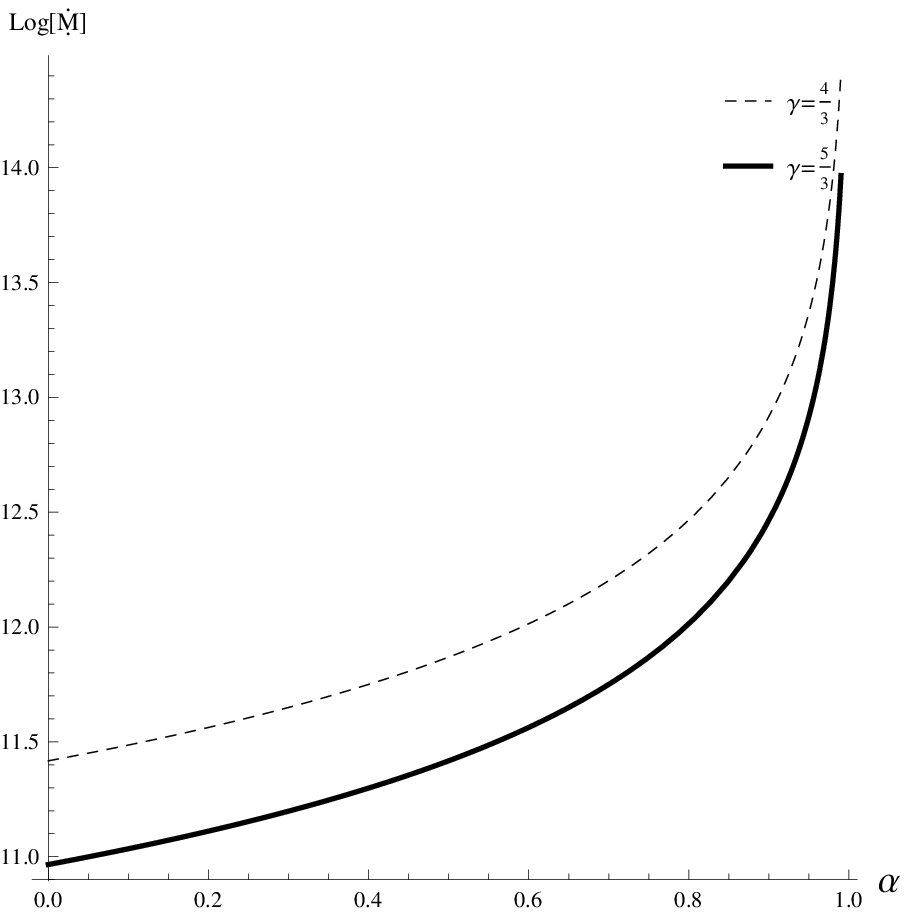}&
\includegraphics[width=7.5 cm, height=6 cm]{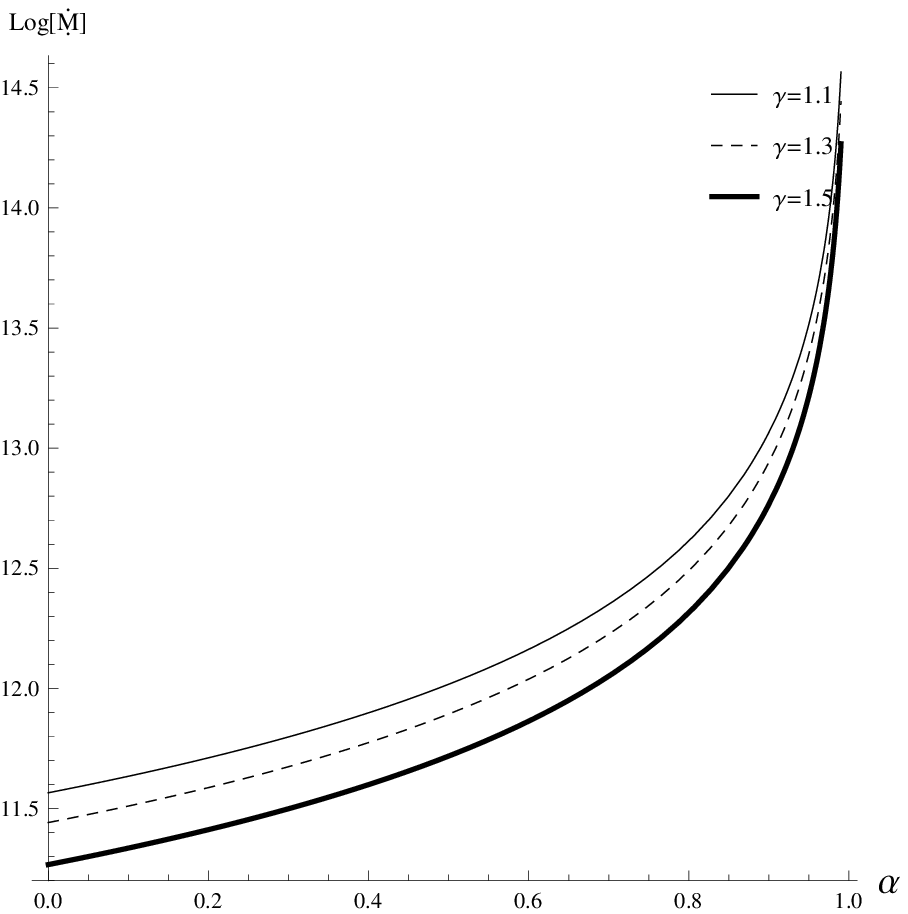}
\\
\hline
\end{tabular}
\caption{\label{AH} Plots showing the logarithm of the accretion rate $\dot{M}$ as a
function of $\alpha$ for different values of $\gamma.$}
\label{fig1}
\end{figure}

\end{widetext}

We are now  in  a position to evaluate the accretion rate
$\dot{M}$. Since $\dot{M}$ is independent of $r$,
(\ref{accrate}) must also hold for $r=r_s$. We use the critical
point to determine the Bondi accretion rate $\dot{M} = 4\pi r_{s}^2
m_{b}n_{s}v_{s}$.  By virtue of eqs. (\ref{sonic}), (\ref{sonsound}), (\ref{sonrad})
and (\ref{barden1}) the accretion rate becomes
\begin{equation}
\label{accrate1} \dot{M} = \frac{4\pi}{(1-\alpha)^{3/2}}
\lambda_{s}
M^{2}m_{b}n_{\infty}a_{\infty}^{-3},
\end{equation}
where we have defined the dimensionless accretion eigenvalue
\begin{equation}
 \label{lambda}
 \lambda_{s} =
 \left(\frac{1}{2}\right)^{(\gamma+1)/2(\gamma-1)}\left(\frac{5-3\gamma}{4}\right)^{-(5-3\gamma)/2(\gamma-1)}.
\end{equation}
From  (\ref{accrate1}), it is evident that  the mass accretion in
a string cloud background is increased by the factor
$(1-\alpha)^{-3/2}$, which may result in a  more luminous
source. However, the accretion rate still scales as $\dot{M} \sim
M^2$ which is similar to that of the  Newtonian model \cite{Bondi} as well as
  the relativistic case \cite{Michel,stbook}.  In the limiting case $\alpha=0$, we obtain the well known
relations derived in  \cite{Michel,stbook} for the Schwarzschild
black hole.  In Fig.~\ref{fig1},  we have plotted the logarithm of
the accretion rate $\dot{M}$ against the string cloud parameter
$\alpha$ for various polytropic indices $\gamma$. Here $\dot{M}$ is
calculated in ergs/sec. We see that $\dot{M}$ increases rapidly with
increasing $\alpha$ ($0\leq \alpha <1$), and interestingly $\dot{M}
\rightarrow \infty$ as $\alpha \rightarrow 1$.

\subsection{Some Numerical Results}
\begin{widetext}

\begin{figure}
\begin{tabular}{c c c}
\includegraphics[width=6 cm, height=5 cm]{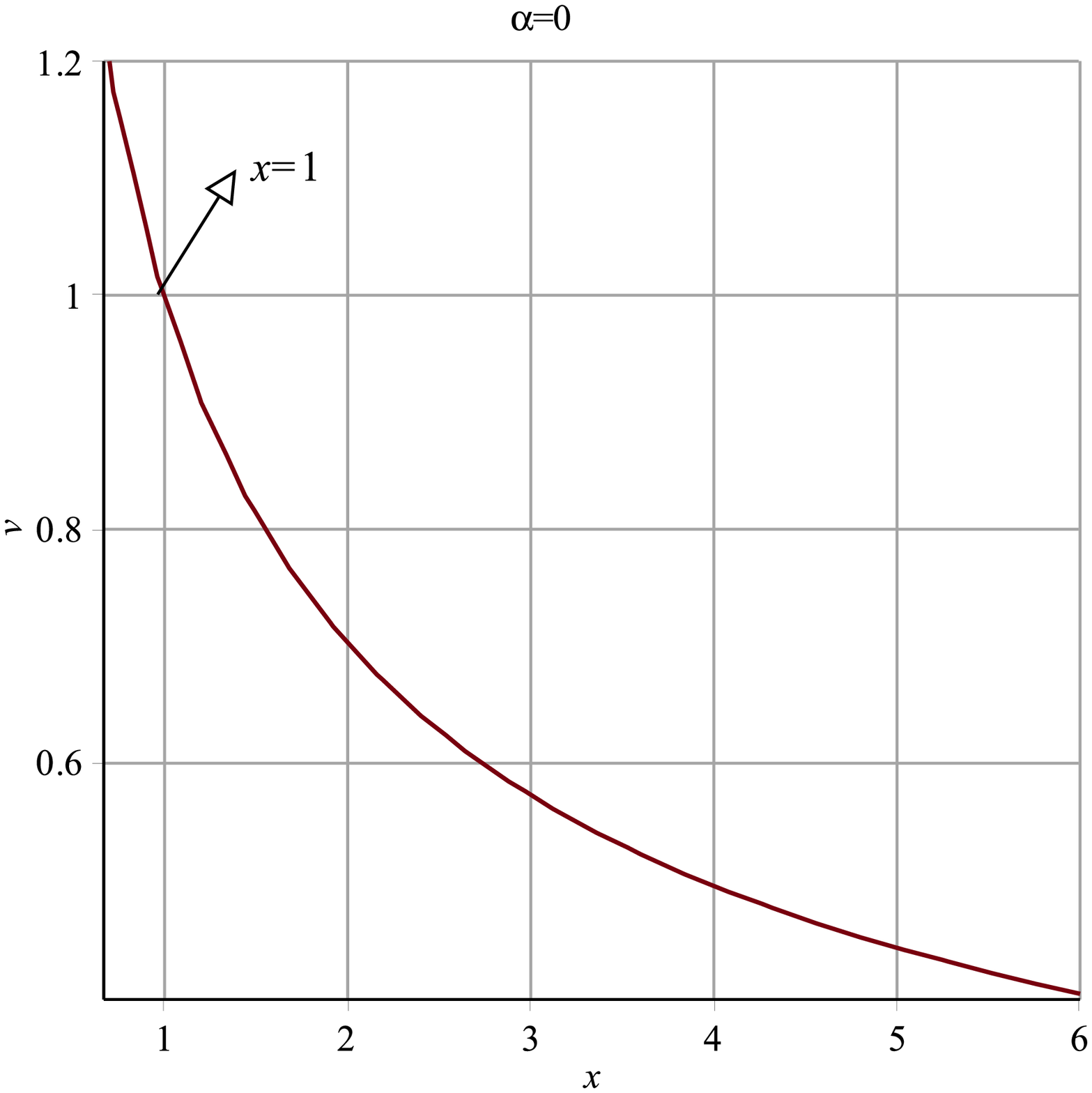}&
\includegraphics[width=6 cm, height=5 cm]{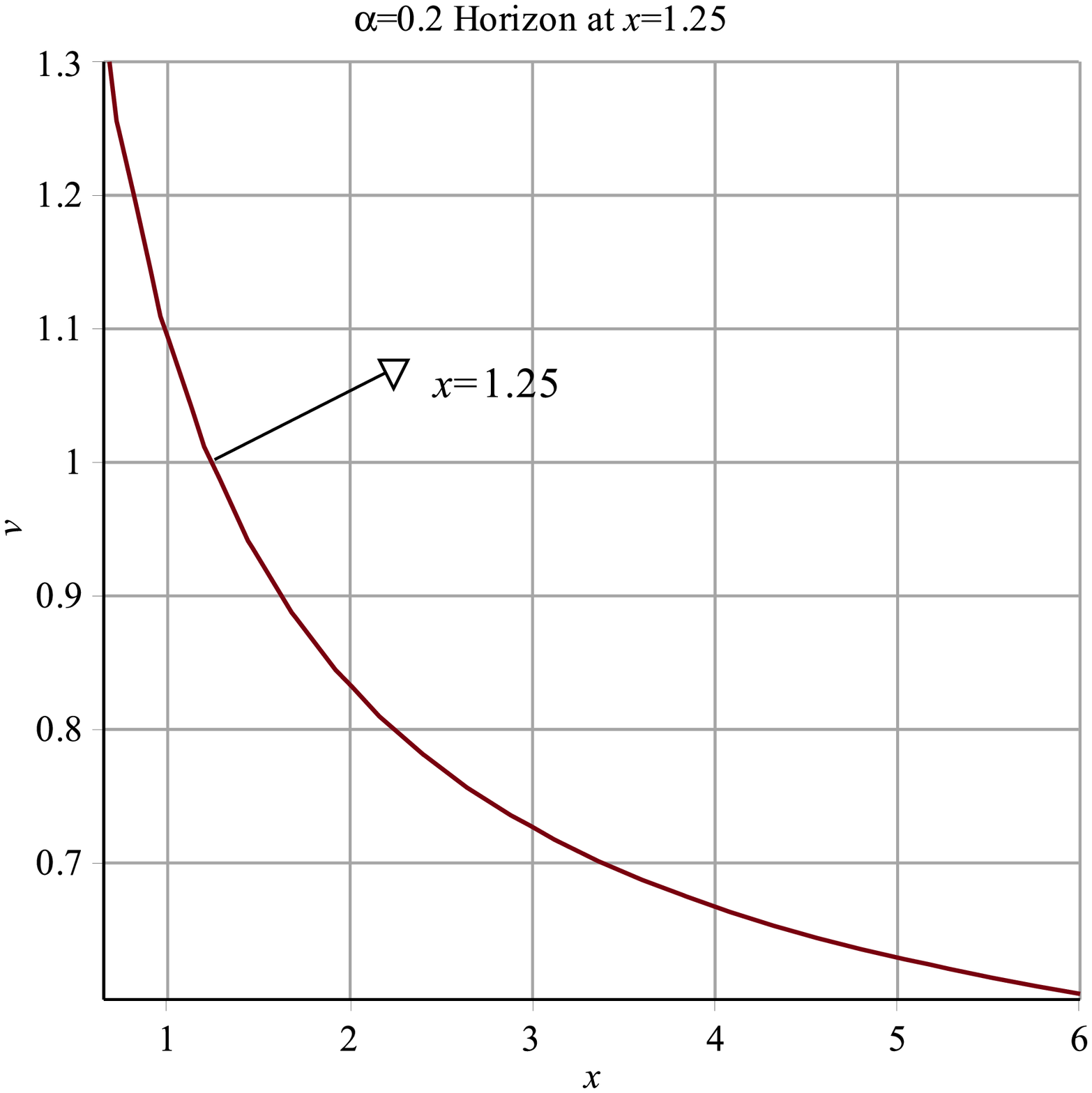}&
\includegraphics[width=6 cm, height=5 cm]{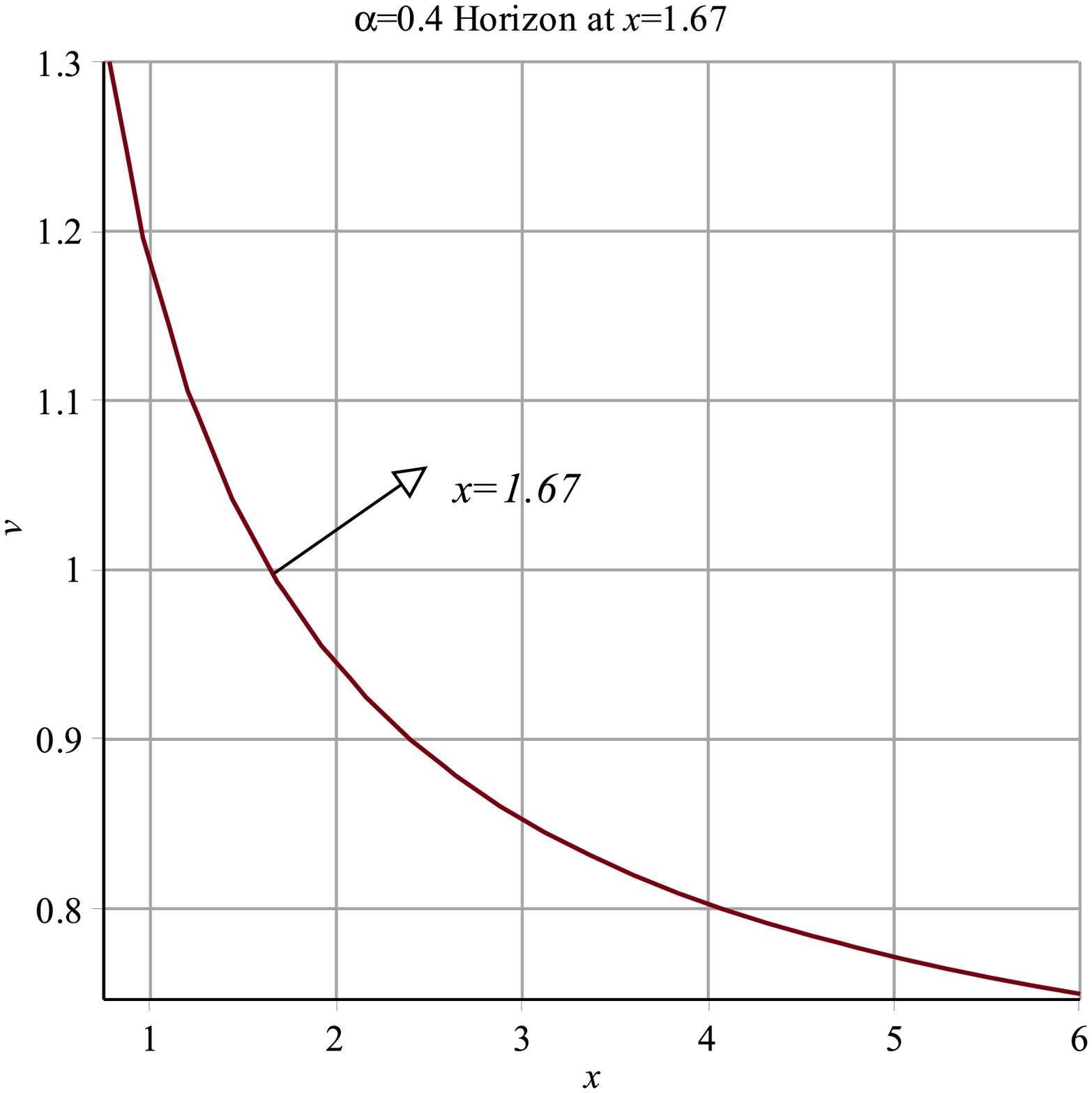} \\
\includegraphics[width=6 cm, height=5 cm]{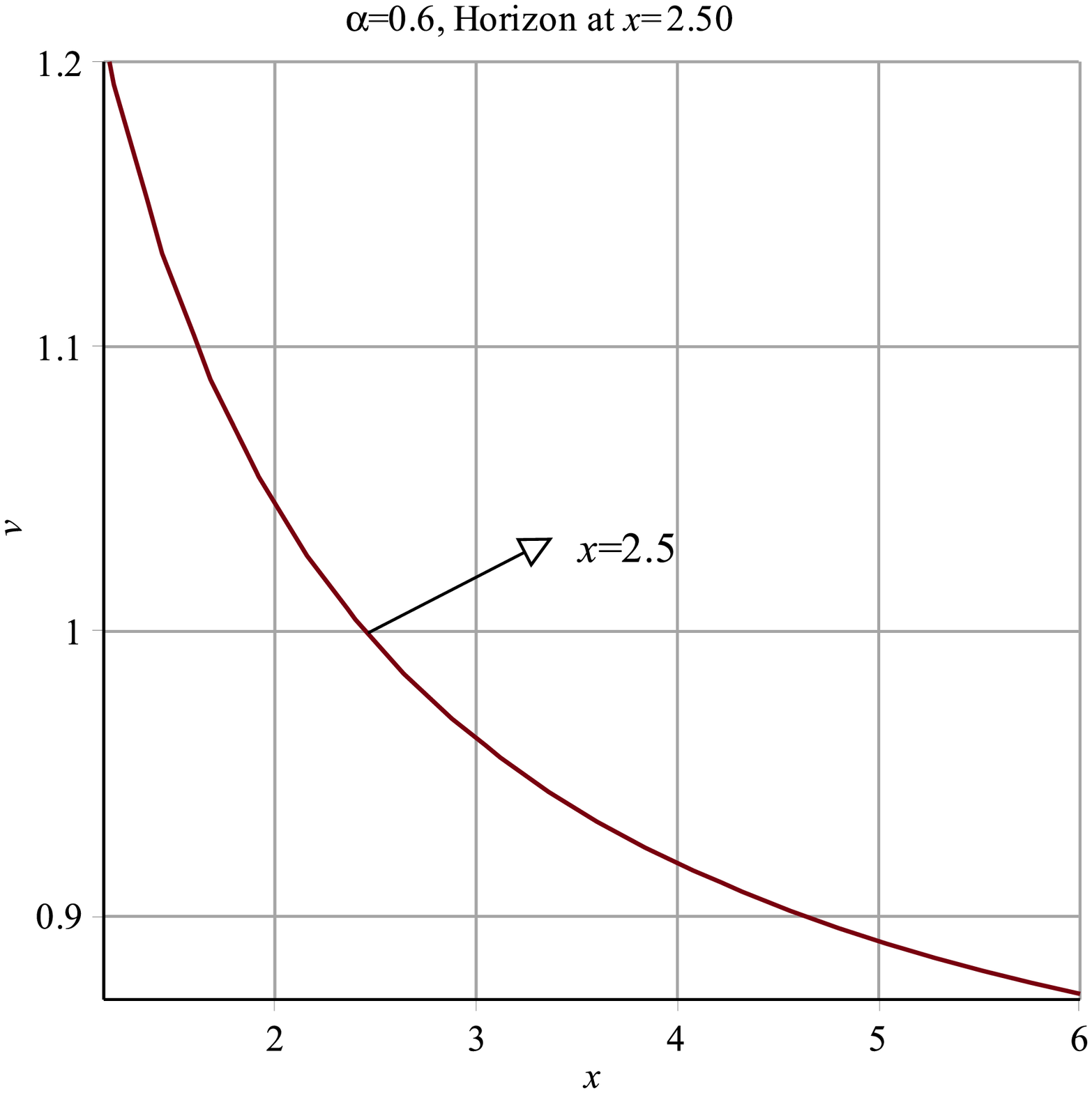}&
\includegraphics[width=6 cm, height=5 cm]{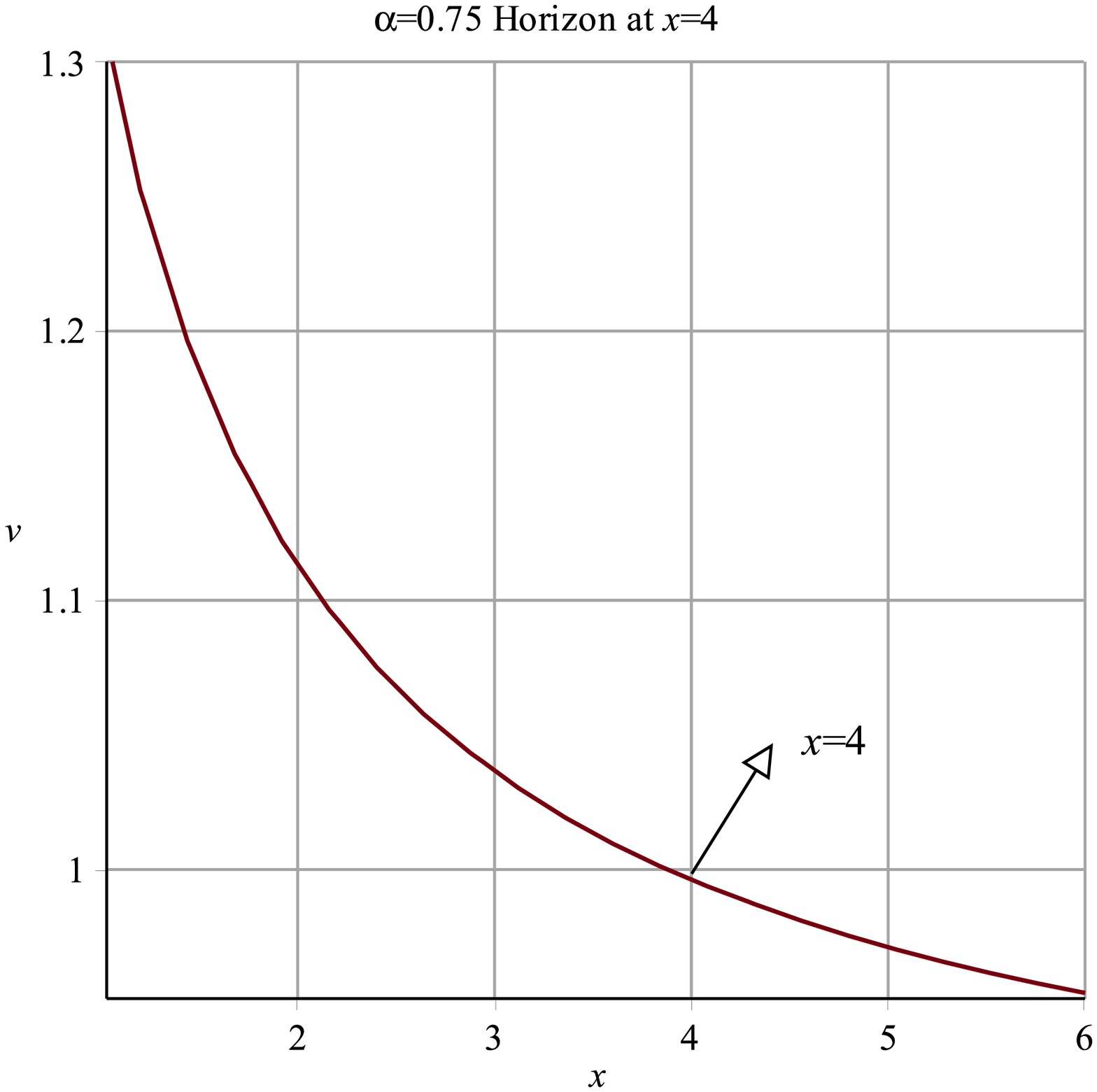}&
\includegraphics[width=6 cm, height=5 cm]{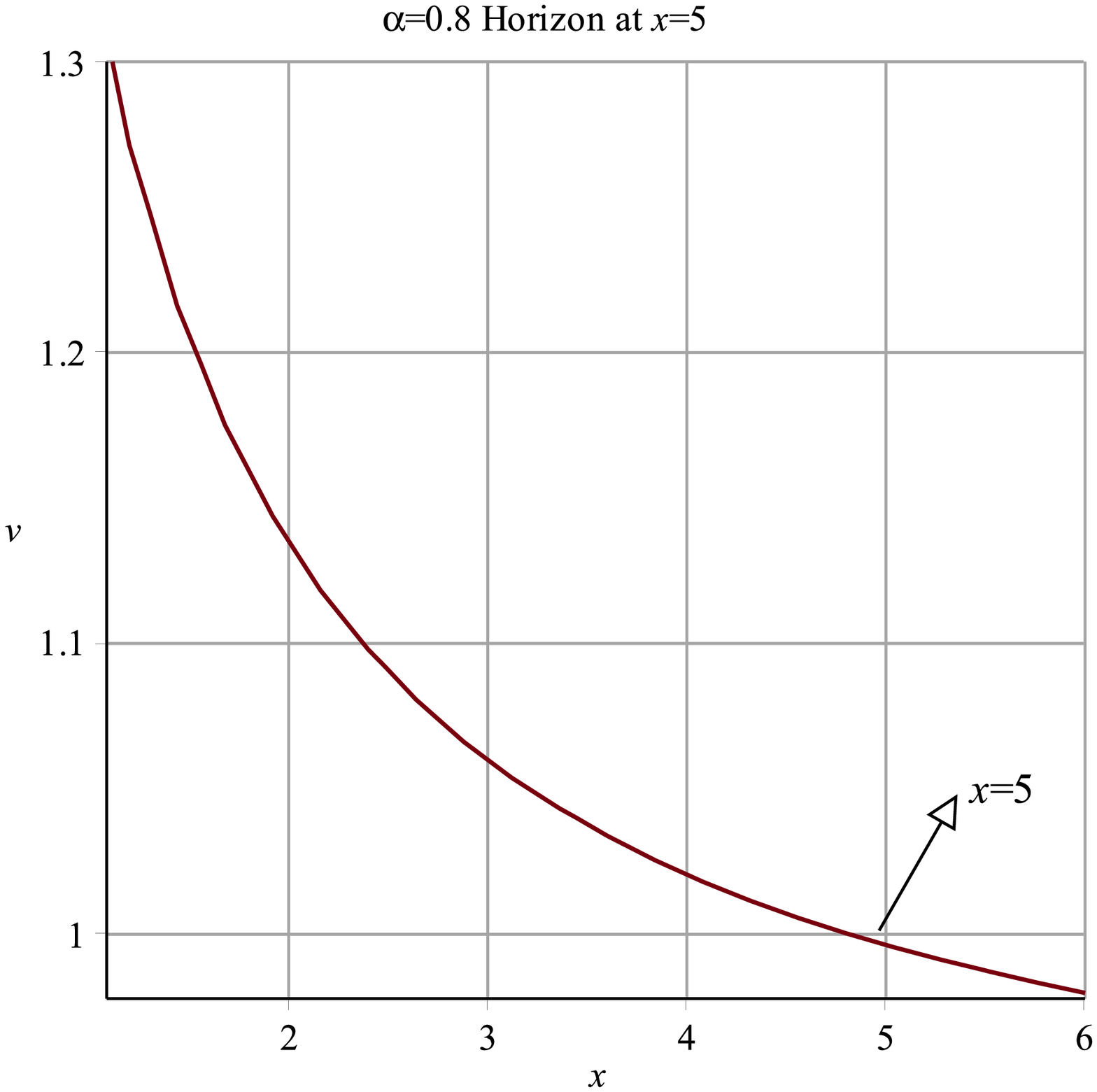} 

\end{tabular}
\caption{The radial velocity profile ($v$)  for a relativistic fluid $\gamma=4/3$ accreting onto the black hole as a function of the dimensionless radius $x=(r/2M)$  for different values of the string cloud parameter $\alpha$.}
\label{fig2}
\end{figure}

\end{widetext}

The radial motion of the relativistic fluid accreting  onto the  black hole in a strings cloud background is governed by (\ref{baryoncons1}) and (\ref{ber1}).  These equations are difficult to solve analytically and we solve them numerically as in ref. \cite{charge1}. We consider only  the case of the relativistic fluid with $\gamma= \frac43$
to study the radial velocity of the flow. Following \cite{charge1}, we introduce dimensionless variables, the radial distance in terms of the
gravitational radius ($x=(r/2M)$) and the particle number density with respect to its value at infinity ($y=n/n_{\infty}$). Now considering $a \ll 1$, Eq.\ (\ref{ber1}) can be rewritten,
in terms of a new variable, as
\begin{align}
\label{num1}
&\left(1 + \frac{a_{\infty}^2}{\gamma - 1 }y^{\gamma-1}\right)^2 \left( 1  - \frac{1-\alpha}{x}-\alpha + v^2 \right)  \nonumber \\
& = \left(1 +  \frac{a_{\infty}^{2}}{\gamma - 1}\right)^2.
\end{align}
On the other hand, using the same notation, the baryon conservation equation (\ref{baryoncons1}) can be recast as
\begin{equation}
\label{num2}
yv=\left(\frac{x_s}{x}\right)^2a_{\infty}\left(\frac{2}{5-3\gamma}\right)^{\gamma+1/2(\gamma-1)}\left(1-\alpha \right)^{1/2},
\end{equation}
where the constant of integration is calculated by applying baryon conservation at the critical point.
Observe that (\ref{num1}) and (\ref{num2})  are corrected equations for the string cloud model and when $\alpha  \rightarrow 0$ we recover the familiar model
of Michel \cite{Michel}. Clearly, the equations (\ref{num1}) and (\ref{num2}) form a nonlinear system of algebraic equations which is
solved numerically for
 the fluid velocity $ v $ given in terms of the velocity of light and $y$. The parameters defining the flow are the sound velocity at infinity $a_{\infty}$, the adiabatic coefficient $\gamma$ and the string cloud parameter $\alpha$. The velocity  profile of the flow as a function
 of the dimensionless variable $x$
for different values of the parameter $\alpha$ is plotted in Fig.~\ref{fig2}. The
 solution is obtained by assuming an asymptotic temperature at infinity of $10^{-9}~m_pc^2/k_B$ for the relativistic case, i.e., $\gamma=\frac43$.

The event horizons for the model are located at $x=1/(1-\alpha)$, and hence the event horizon varies with $\alpha$. Interestingly a string cloud in the background makes
a profound influence on the radial velocity, and the result is strikingly different from the
Schwarzschild case ($\alpha =0$). 
In the familiar Schwarzschild case ($\alpha =0$, Fig.~\ref{fig2}), we note that the flow speed of the accreting gas
 crosses the event horizon at the speed of light. 
 This feature is consistent with the treatment of de Freitas \cite{charge1}
  who considered relativistic accretion
  onto a charged black hole. 
 The critical radius is far away from the event horizon ($x_{c}=1.25 \times 10^{8}$) where the flow velocity is much less the value at the event horizon. To conserve space we have plotted velocity profile for $\gamma = 4/3$, as  the radial velocity $v$ for other values of $\gamma$ have similar profiles.  The velocity profile are plotted for some specific values of string cloud parameter $\alpha=0,\; 0.2,\;0.4,\;0.6,\; 0.75$ and $0.8$, respectively for which the event horizons are located at $r=1,1.25,1.67,2.5,4$ and $5$.  It is clear from Fig.~\ref{fig2}, the fluid  always crosses event horizon with the velocity of light for all values of $\alpha$.

\begin{widetext}
 
 \begin{figure}
 \begin{tabular}{c c c}
 \includegraphics[width=6 cm, height=5 cm]{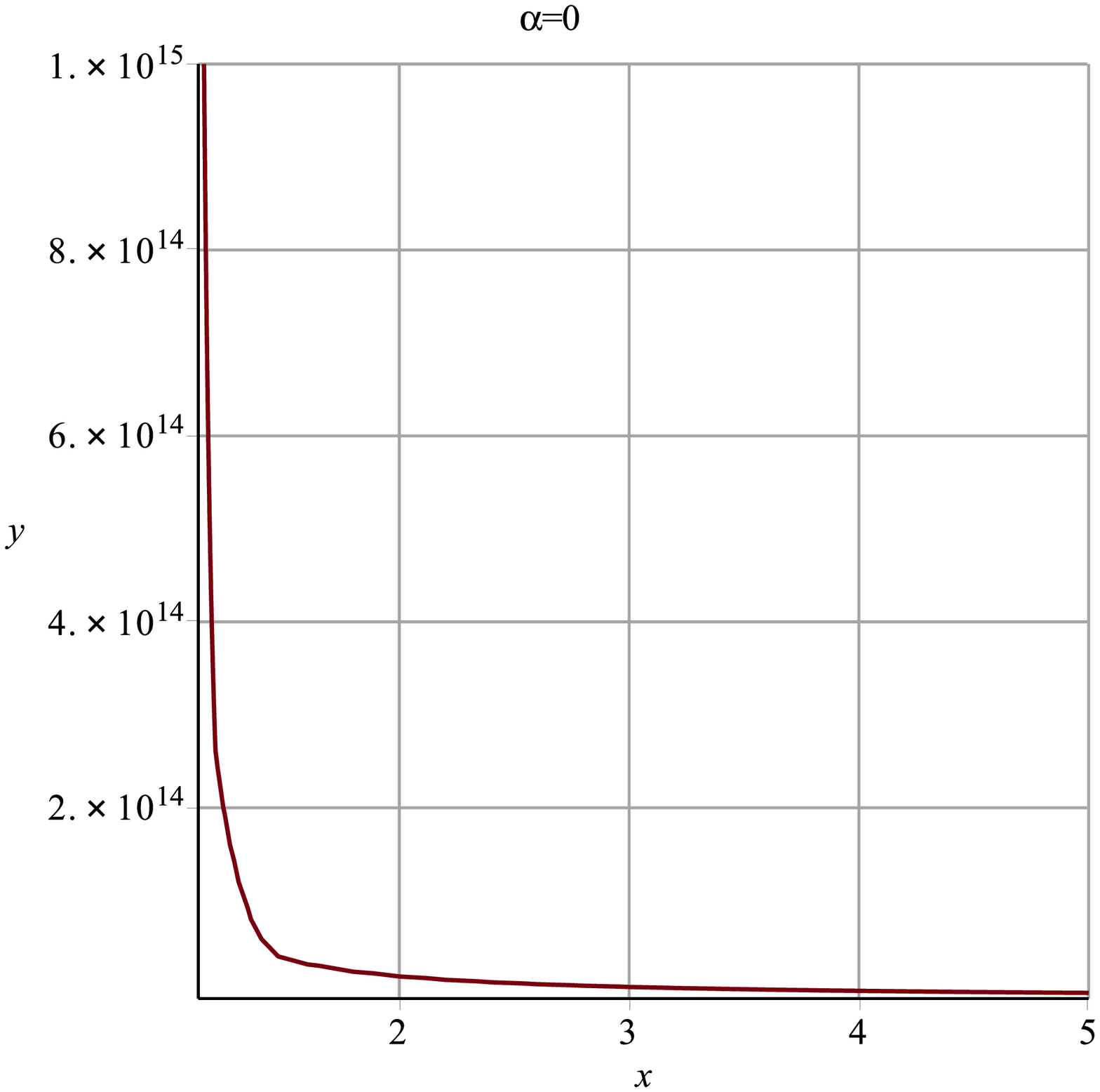}&
 \includegraphics[width=6 cm, height=5 cm]{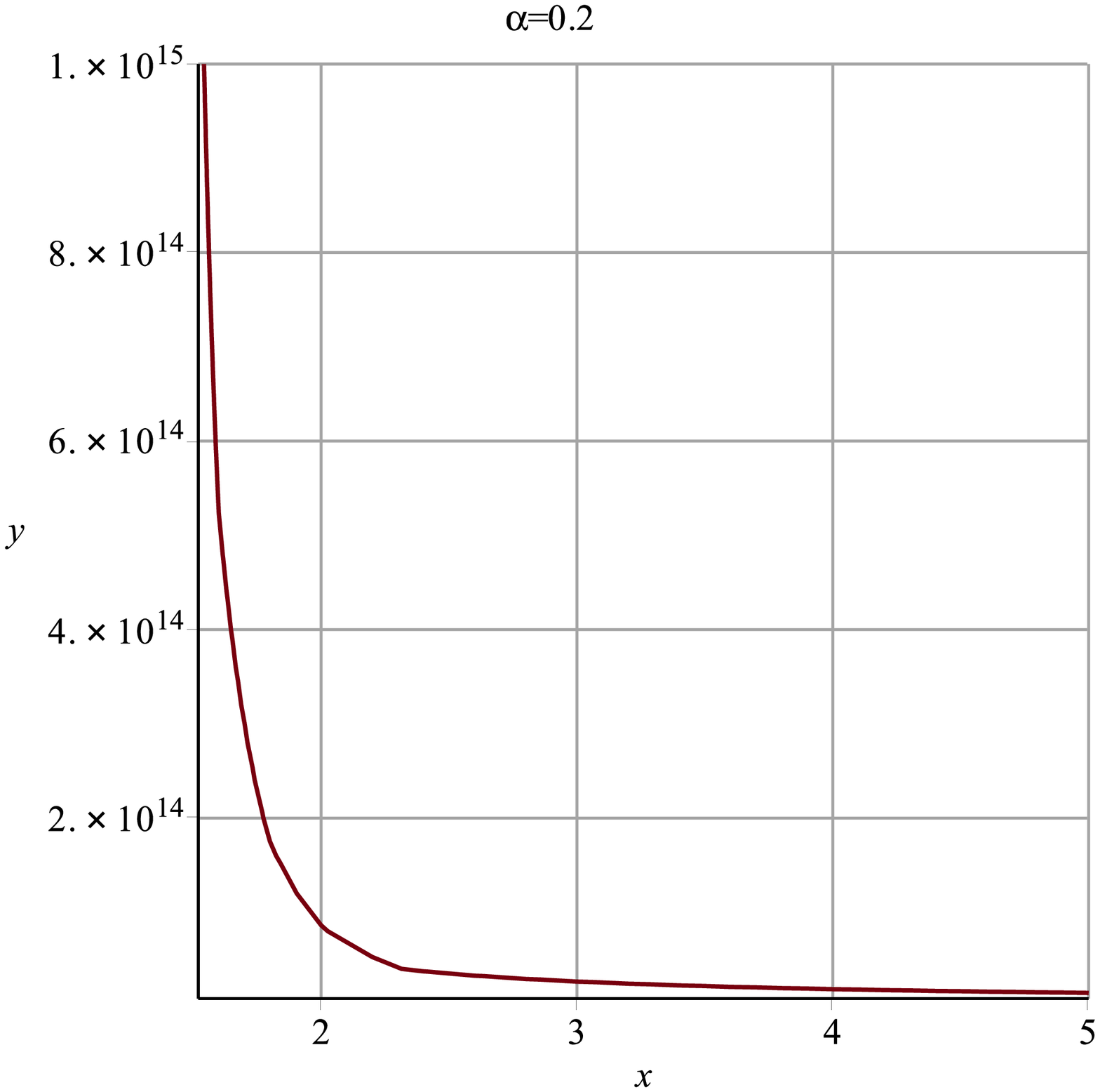}&
 \includegraphics[width=6 cm, height=5 cm]{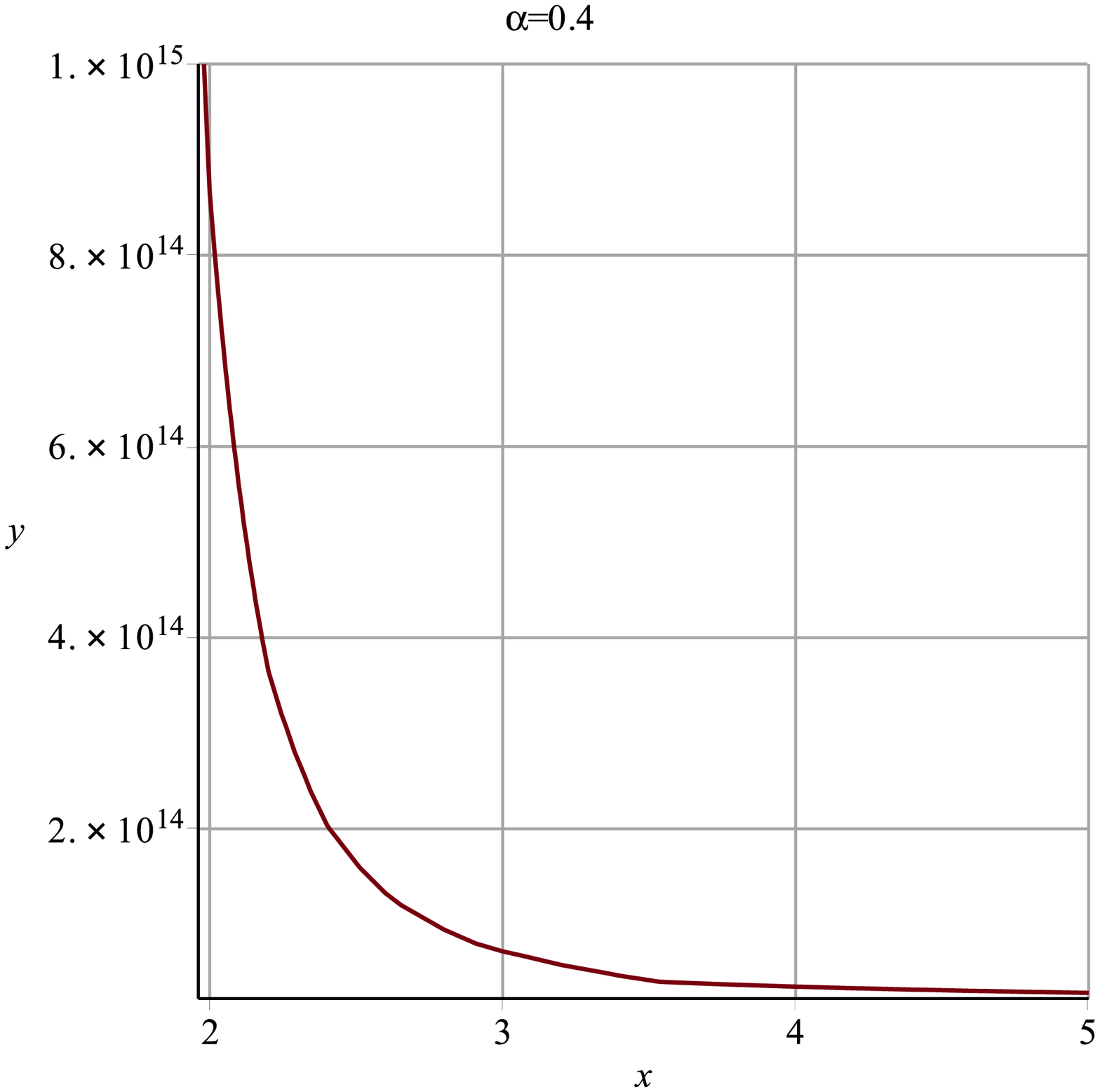} \\
 \includegraphics[width=6 cm, height=5 cm]{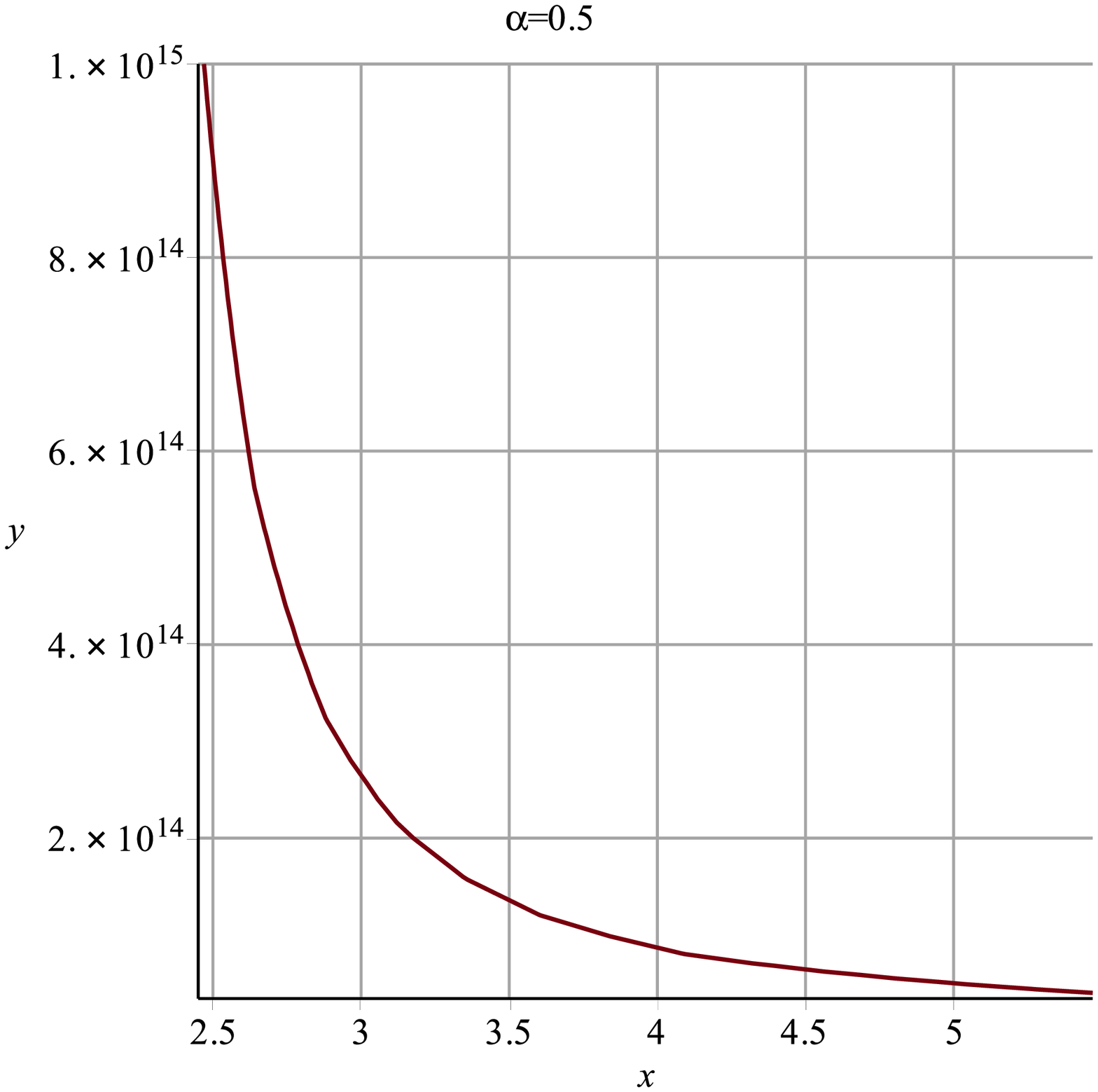}&
 \includegraphics[width=6 cm, height=5 cm]{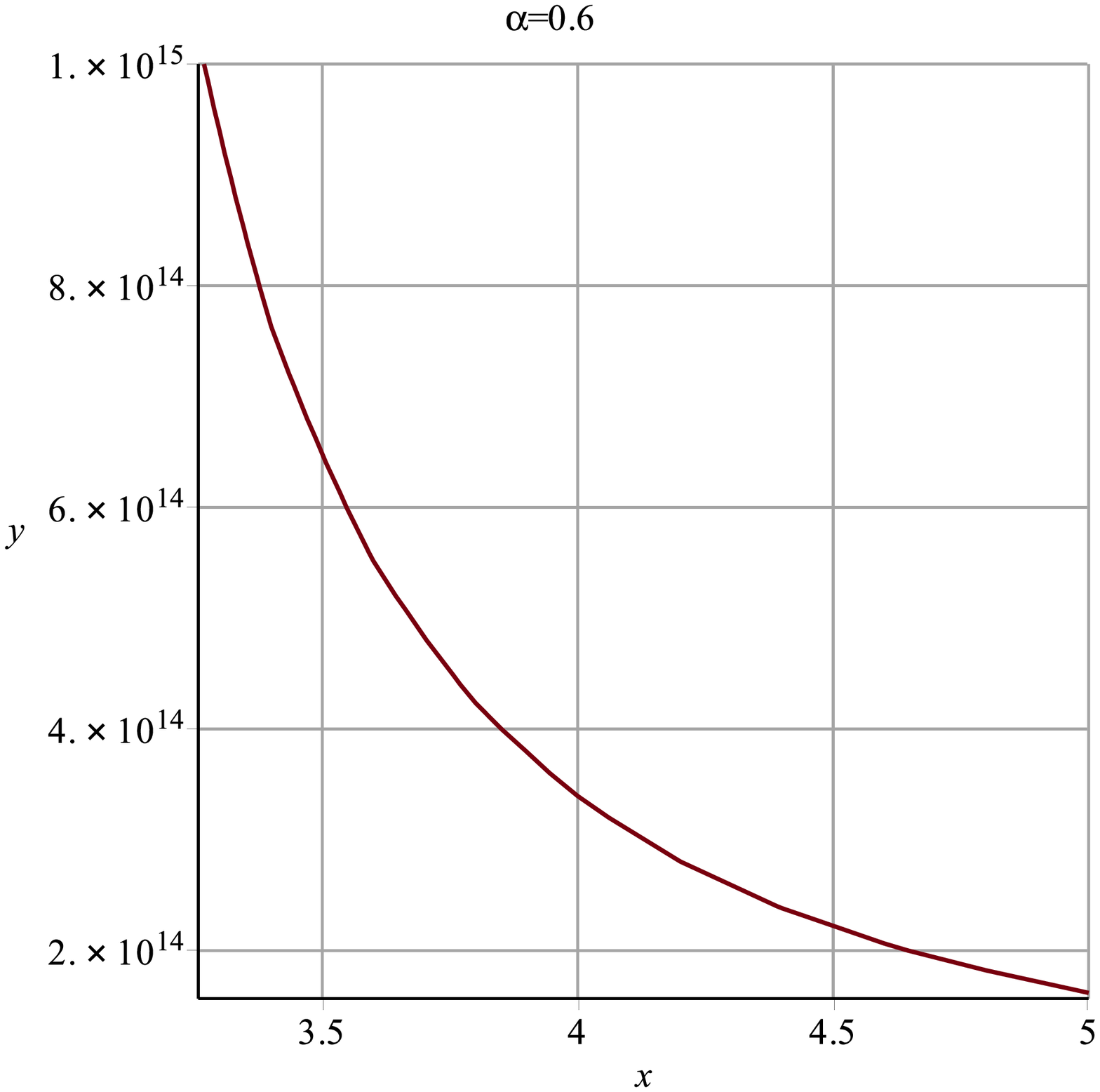}&
 \includegraphics[width=6 cm, height=5 cm]{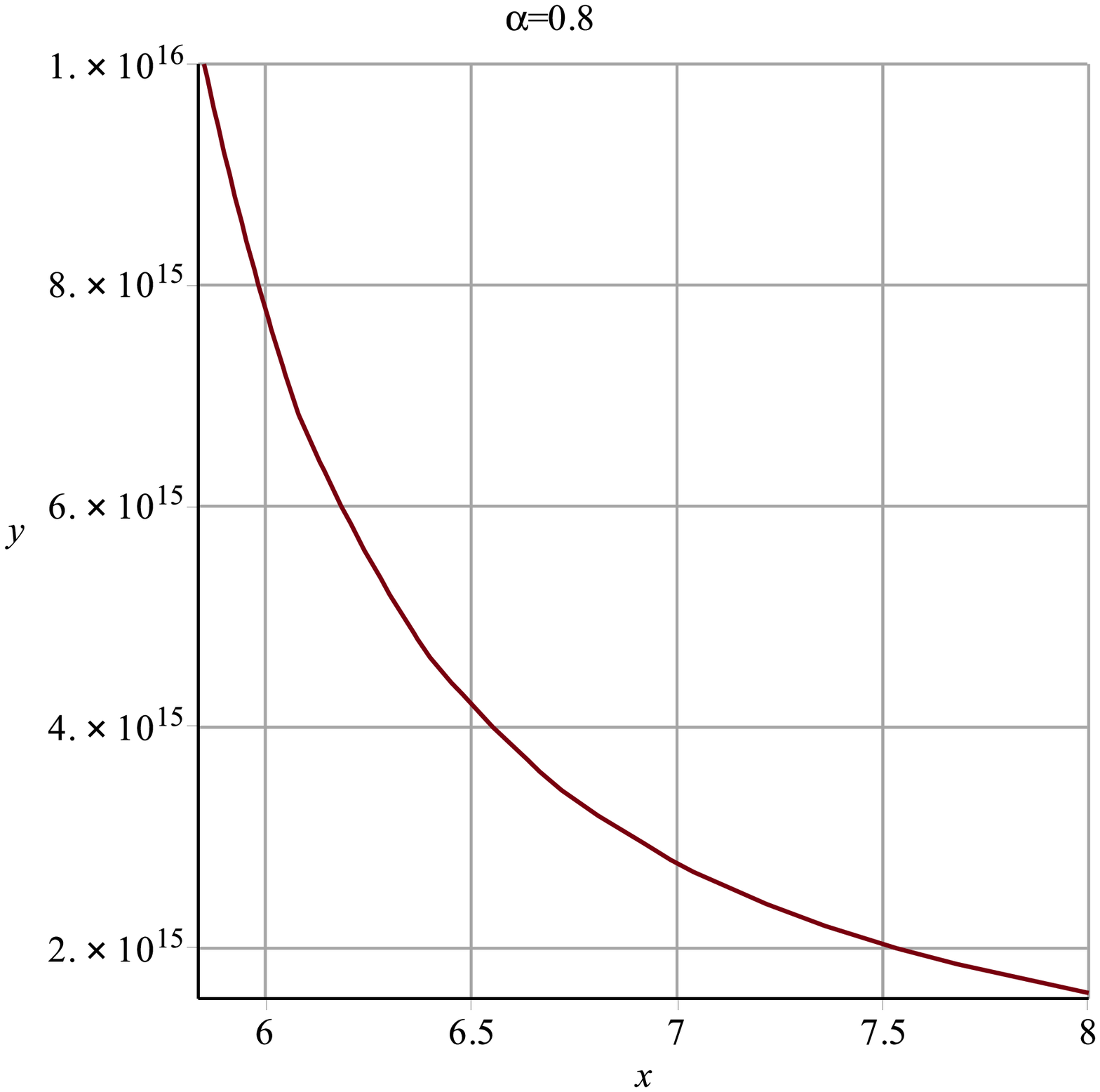} 
 
 \end{tabular}
 \caption{The  compression factor ($y$) for a relativistic fluid $\gamma=4/3$ accreting onto the black hole as a function of the dimensionless radius $x=(r/2M)$  for  different values of the string cloud parameter  $\alpha$.}
 \label{fig3}
 \end{figure}
 
 \end{widetext}

 We have also plotted the compression ratio $y$ as a function of radial coordinate for a relativistic accreting gas with $\gamma = \frac43$ in Fig.~\ref{fig3} for different values of string cloud parameter. more specifically for $\alpha=0,\; 0.2,\;0.4,\;0.6,\; 0.75$ and $0.8$. This graph
 shows that the compression factor profiles also affected by a change in the strings cloud parameter $\alpha$ this is in contrast to analogous
 compression factor of an accreting charged black hole \cite{charge1}.   The compression ratio for black hole in string cloud background increases with increase in $\alpha$. In general, it may attains the value of the order of $10^{14}-10^{16}$. 
 
\section{Asymptotic behaviour}

In the last section, we found that the accretion rate at some sonic point $r=r_s$  far away from event horizon, i.e., $r_s \gg 2M$ is not influenced by nonlinear gravity.
Next we estimate the flow characteristics for $r_H < r
\ll r_s$ and at the event horizon $r = r_H$.

\subsection{Sub-Bondi radius $r_H < r \ll r_s$}

 At distances below the Bondi radius the gas is supersonic so that  $v>a$ when $r_H < r \ll r_s$. From (\ref{ber1}) we find the upper bound on the radial dependence of the gas velocity
\begin{equation}
\label{supsonic}
v^2 \approx \frac{2M}{r}, \; \qquad \gamma \neq \frac{5}{3}.
\end{equation}
We can now estimate the gas compression on these scales. With the help of (\ref{accrate}), (\ref{accrate1}) and (\ref{supsonic}) we obtain
\begin{equation}
\label{gascomp}
\frac{n(r)}{n_{\infty}} \approx \frac{\lambda_{s}}{\sqrt{2}(1-\alpha)^2} \left( \frac{M}{a^2_{\infty}r} \right)^{3/2}.
\end{equation}
For a Maxwell-Boltzmann gas, $p = nk_{B}T$, we generate the adiabatic temperature profile
\begin{equation}
\label{temp}
\frac{T(r)}{T_{\infty}} = \left(\frac{n(r)}{n_{\infty}}\right)^{\gamma-1}\approx \left[\frac{\lambda_{s}}{\sqrt{2}(1-\alpha)^2} \left( \frac{M}{a^2_{\infty}r} \right)^{3/2}\right]^{\gamma-1},
\end{equation}
using (\ref{eos}) and (\ref{gascomp}).

\subsection{Event horizon}

At the event horizon we have $r = r_H = 2M/(1-\alpha)$.  As the flow is supersonic since we are well below the Bondi radius,
it is reasonable to assume that the fluid velocity is approximated  by $v^2 \approx \frac{2M}{r}$. At $r_H$, $v_{H}^{2} \equiv v^2(r_H) \approx 1-\alpha $, i.e., the flow speed at the horizon is always less than the speed of light. Therefore, using the fact $M/r_H=(1-\alpha)/2$, we obtain the gas compression at the event horizon from (\ref{gascomp}):
\begin{equation}
\label{horcomp}
\frac{n_H}{n_{\infty}} \approx \frac{\lambda_{s}}{4(1-\alpha)^{1/2}}\left( \frac{c}{a_\infty} \right)^3.
\end{equation}
Again assuming the presence of a  Maxwell-Boltzmann gas, $p = nk_BT$, we find the adiabatic temperature profile at the event horizon using (\ref{temp}) and the horizon assumption:
\begin{equation}
\label{hortemp}
\frac{T_H}{T_{\infty}} \approx \left[ \frac{\lambda_{s}}{4(1-\alpha)^{1/2}}\left( \frac{c}{a_\infty} \right)^3\right]^{\gamma -1},
\end{equation}
where, following \cite{stbook}, we have re-introduced the speed of light $c$ in the above expressions. The limit $\alpha\rightarrow 0$ in the above equation gives us the corresponding result of  accretion of the fluid onto the Schwarzschild black hole \cite{stbook}.

\section{Conclusions}
Historically the accretion problem with a polytropic
equation of state  was addressed by Bondi \cite{Bondi}. He showed 
 that subsonic flow far from a black hole will inevitably
become supersonic, and that the requirement of a smooth traversal of
the sonic surface uniquely specifies the accretion rate as a
function of two thermodynamic variables, namely the density and
temperature of the gas at infinity.  The relativistic version of the
same problem was solved by Michel \cite{Michel} twenty years later,
after the discovery of celestial X-ray sources. He showed that
accretion onto the black hole should be transonic.
Accretion onto
compact objects such as black holes and neutron stars is the most
efficient method of releasing energy;  up to 40 percent of the
rest-mass energy of the matter accreting on the black hole is
liberated.  Recent developments in the theory of
accretion are significant steps toward understanding various
astronomical sources that are believed to be powered by the
accretion onto black holes. Spherical accretion onto a black hole is
generally specified by the mass accretion rate $\dot{M}$ which is a
key parameter, and there is evidence that a higher accretion rate
can provide higher luminosity values. In view of this, we analyzed
the steady state and spherical accretion of a fluid onto the
Schwarzschild black hole in a string cloud background.  We
determined exact expressions for the mass accretion rate at the
critical radius.  It turns out that this quantity is modified so
that $ \dot{M} \approx {M^2}/{(1-\alpha)^{3/2}} $ with $r_s \approx
{M}/{(1-\alpha)} $.  Thus the accretion rate by the black hole in a
string cloud background is higher than that for a Schwarzschild
black hole.  Thus the parameter $\alpha$ can be introduced in the
problem of accretion onto black hole to extend the work of Michel
\cite{Michel}, and this quantity determines the accretion rate and
other flow parameters. In principle, the accretion rate and other
parameters still have same characteristics as in the Schwarzschild
black hole;
 in this sense we may conclude that the familiar steady state spherical accretion solution onto the Schwarzschild black hole is
 stable.
 In the limit $\alpha \rightarrow 0$, our results reduce exactly to those obtained
in \cite{Michel,stbook} for the standard Schwarzschild black hole.

We can attempt to work out the effect of string cloud background on
the luminosity, the frequency spectrum and the energy conversion
efficiency of the accretion flow.  It is possible to deviate from
spherical symmetry, e.g., include rotation, which may lead to a
higher accretion rate. These and other related issues are currently
under investigation.

\section*{Acknowledgements}

A.G. thanks the N.R.F. and U.K.Z.N. for financial support. A.G. is also
grateful to the facilities provided at J.M.I., New Delhi, where part
of the work has been done. S.G.G.  thanks the
University Grant Commission (UGC) for the major research project
grant F. NO. 39-459/2010 (SR) and to M. Sami for useful discussions. S.D.M. acknowledges that this work is
based upon research supported by the South African Research Chair
Initiative of the Department of Science and Technology and the
National Research Foundation.


\begin{thebibliography}{99}
\bibitem{Chakrabarti:1996cc} S.K. Chakrabarti,   Phys. Rept.  {\bf 266}, 229 (1996).

\bibitem{Bondi} H. Bondi, Mon. Not. Roy. Astron. Soc. {\bf 112}, 195 (1952).

\bibitem{Michel} F.C. Michel,
  Astrophys. Space Sci. {\bf 15}, 153 (1972).

\bibitem{keylist} B. J. Carr and S.W. Hawking, Mon. Not. R. Astron. Soc. {\bf168}, 399
(1974);  M.C. Begelman, Astron. Astrophys. {\bf 70}, 583 (1978); D.
Ray, Astron. Astrophys. {\bf 82}, 368 (1980);   K.S. Thorne, R.A.
Flammang and A.N. Zytkow, Mon. Not. R. Astron. Soc. {\bf 194}, 475
(1981);  E. Bettwieser and W. Glatzel, Astron. Astrophys. {\bf 94}, 306
(1981);  K.M. Chang, Astron. Astrophys. {\bf 142}, 212 (1985);   U.S.
Pandey, Astrophys. Space Sci. {\bf 136}, 195 (1987).

\bibitem{stbook}   S.L. Shapiro and S.A. Teukolsky,
   \textit{Black Holes, White Dwarfs and Neutron Stars: The Physics of Compact Objects} (Wiley, New York, 1983).

\bibitem{shap73a} S.L. Shapiro, Astrophys. J. {\bf 180}, 531 (1973).

\bibitem{shap73b} S.L. Shapiro, Astrophys. J. {\bf 185}, 69 (1973).

\bibitem{shap74} S.L. Shapiro, Astrophys. J. {\bf 189}, 343 (1974).

\bibitem{blum} G.R. Blumenthal and W.G. Mathews, Astrophys. J. {\bf 203}, 714 (1976).

\bibitem{brink} W. Brinkmann, Astron.  Astrophys. {\bf 85}, 146 (1980).

\bibitem{malec} E. Malec, Phys. Rev. D {\bf 60}, 104043 (1999).

\bibitem{perfect} E. Babichev, V. Dokuchaev and Y. Eroshenko,  Phys. Rev. Lett. {\bf 93},
021102 (2004);   E. Babichev, V. Dokuchaev and Y. Eroshenko, AIP
Conf. Proc. {\bf 861}, 554 (2006).

\bibitem{perfect1} E. Babichev, V. Dokuchaev  and Y. Eroshenko, J. Exp. Theor. Phys.
{\bf 100}, 528 (2005); E. Babichev, V. Dokuchaev and Y. Eroshenko, Zh. Eksp. Teor.
Fiz. {\bf 127}, 597 (2004).

\bibitem{charge1} J. A. de Freitas Pacheco, Journal of Thermodynamics {\bf 2012}, 791870 (2012).

\bibitem{charge} E. Babichev, S. Chernov, V. Dokuchaev and Y. Eroshenko,  J. Exp.
Theor. Phys. {\bf 112}, 784 (2011).


\bibitem{Jamil}  M. Jamil, M.A. Rashid and A. Qadir,
  Eur. Phys. J. C {\bf 58}, 325 (2008).

\bibitem{Sharif11}  M. Sharif and G. Abbas,
  Mod. Phys. Lett. A {\bf 26}, 1731 (2011).

\bibitem{Sharif12}  M. Sharif and G. Abbas,
  Chin.  Phys. Lett.  {\bf 29}, 010401 (2012).
  

\bibitem{Le1} P. Letelier,  Phys. Rev. D {\bf 20}, 1294 (1979).

\bibitem{Synge}  J.L. Synge, \textit{Relativity: The General theory}
  (North-Holland, Amsterdam, 1960).

\bibitem{Kibble}  T. Kibble, J. Phys. A {\bf 9}, 1387 (1976).

\bibitem{Vilenkin} A. Vilenkin, Phys. Rev. Lett. {\bf 46}, 1169 (1981).

\bibitem{dn} 
  D.~Mitchell and N.~Turok,
  Nucl.\ Phys.\ B {\bf 294}, 1138 (1987).
  
\bibitem{Planks} 
  P.A.R. Ade {\it et al.}  [Planck Collaboration],
  arXiv:1303.5085 [astro-ph.CO].



\bibitem{as} A.~Sen,
 ``Developments in superstring theory,''
  In *Vancouver 1998, High energy physics, vol. 1* 371-381
  [hep-ph/9810356].
  
\bibitem{fl} 
  F.~Larsen,
  Phys.\ Rev.\ D {\bf 56}, 1005 (1997).
  

\bibitem{Strominger:1996sh} 
  A.~Strominger and C.~Vafa,
  Phys.\ Lett.\ B {\bf 379}, 99 (1996)


\bibitem{pv} 
  R.~Parthasarathy and K.~S.~Viswanathan,
  Phys.\ Lett.\ B {\bf 400}, 27 (1997).

\bibitem{sgr}  
  S.~H.~Mazharimousavi, O.~Gurtug and M.~Halilsoy,
  Class.\ Quant.\ Grav.\  {\bf 27}, 205022 (2010).

\bibitem{sgb}  
  E.~Herscovich and M.~G.~Richarte,
  Phys.\ Lett.\ B {\bf 689}, 192 (2010).

\bibitem{sll}  
  S.~G.~Ghosh and S.~D.~Maharaj, 
  Phys. Rev. D {\bf 89} 84027 (2014).

\bibitem{gk}  
  E.~N.~Glass and J.~P.~Krisch,
  Phys.\ Rev.\ D {\bf 57}, 5945 (1998);
  J.\ Math.\ Phys.\  {\bf 40}, 4056 (1999);
  Class.\ Quant.\ Grav.\  {\bf 16}, 1175 (1999).

\bibitem{John} A.J. John, S.G. Ghosh and S.D. Maharaj, Phys. Rev. D {\bf 88}, 104005 (2013).


\end{thebibliography}
\end{document}